\pdfoutput=1
\documentclass[10pt,letterpaper]{article}

\usepackage{scbench-preprint}

\preprintKicker{bioRxiv preprint}
\preprintCategory{Systems Biology}
\preprintDate{May 2026}
\preprintTitle{scBench-Long:\\Verifiable Benchmarking\\of Long-Horizon\\Single-Cell Biology}
\preprintSubtitle{Evaluating whether AI agents can recover complex scientific conclusions from raw single-cell biology data}
\preprintAuthors{Ian Diks\preprintSup{1,*}, Zhen Yang\preprintSup{1,*}, Arjun Banerjee\preprintSup{1}, Tim Proctor\preprintSup{1}, Kenny Workman\preprintSup{1}}
\preprintAffiliations{\preprintSup{1}LatchBio, San Francisco, CA, USA\\\preprintSup{*}Equal first authors}
\preprintCorrespondence{kenny@latch.bio}

\begin{document}

\PreprintHeader

\begin{PreprintAbstract}

Single-cell studies require analysts to convert raw measurements into specific
biological claims through multi-step workflows and integration of metadata,
assay context, and auxiliary evidence. Existing AI-biology benchmarks largely
measure broad knowledge, executable workflows, or local analysis steps. We
introduce scBench-Long, a benchmark for long-horizon single-cell biology in
which agents must recover scientific conclusions from raw or near-raw data
without prescribed methods. The benchmark contains 21 evaluations spanning
melanoma CD8 T-cell reactivity, CD8 RNA+ATAC regulatory inference, human--monkey
chimera development, KRAS-driven lung tumor aging, and lethal COVID-19 lung
pathology. Tasks cover paired scRNA/TCR sequencing, RNA and chromatin profiling,
cross-species transcriptomics, combinatorial scRNA-seq, single-nucleus RNA-seq,
immune repertoires, ortholog maps, ligand--receptor resources, and validation
evidence. Candidate claims are reproduced, reviewed, and converted into
controlled answer vocabularies with deterministic grading and trajectory
rubrics. Across 1,068 completed trajectories, the strongest model--harness pair
passes 16/63 runs (25.4\%). scBench-Long evaluates whether agents can move beyond
local analysis steps and make complex scientific claims that are supported by
single-cell data.
\end{PreprintAbstract}

\vfill

\clearpage

\section*{Topline Benchmark Performance}
{\fontsize{9.1}{12.8}\selectfont
We ran the current evaluation set across frontier model families and agent harnesses. Passing requires
exact recovery of the graded structured answer for an evaluation attempt. Present systems show low
but nonzero success rates: the strongest model--harness pair passes 16/63 completed runs across a
1,068-trajectory benchmark matrix, while most pairs solve only a small minority of tasks. This is
consistent with the benchmark targeting long-horizon scientific workflow control rather than
isolated tool execution.\par}

\begin{center}
  \includegraphics[width=0.94\textwidth]{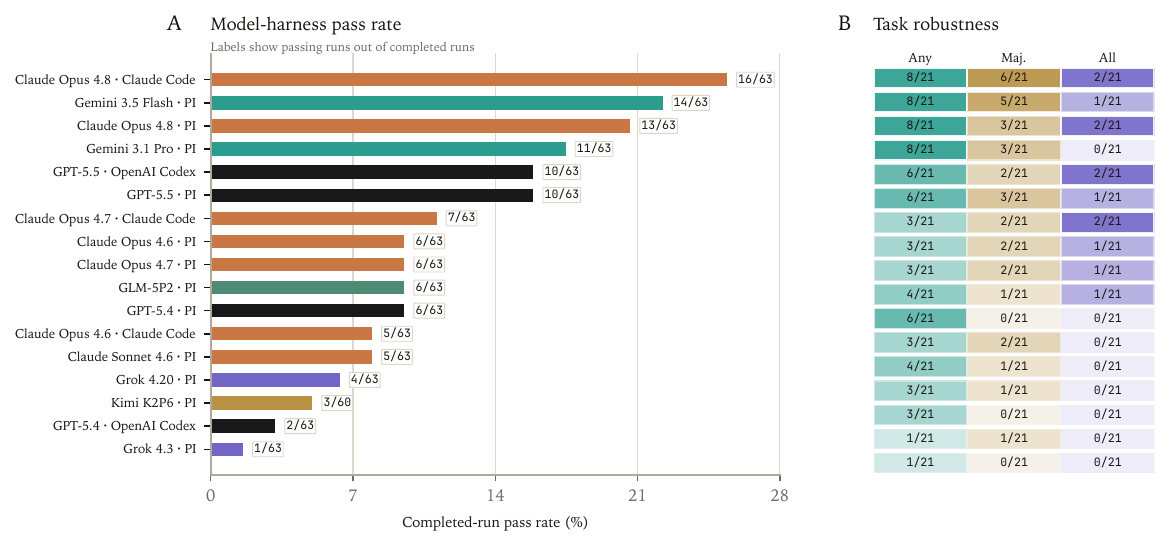}

  \captionof{figure}{\textbf{Topline benchmark performance.}
  Completed-run pass rates and task-level robustness counts across 21 final evaluations and
  17 model--harness pairs. Wilson confidence intervals are reported in
  Table~\ref{tab:model-results} rather than plotted because runs are grouped within a shared
  evaluation set.}
  \label{fig:topline-performance}
\end{center}

\vspace{0.02in}

\begin{center}
  \includegraphics[width=0.94\textwidth]{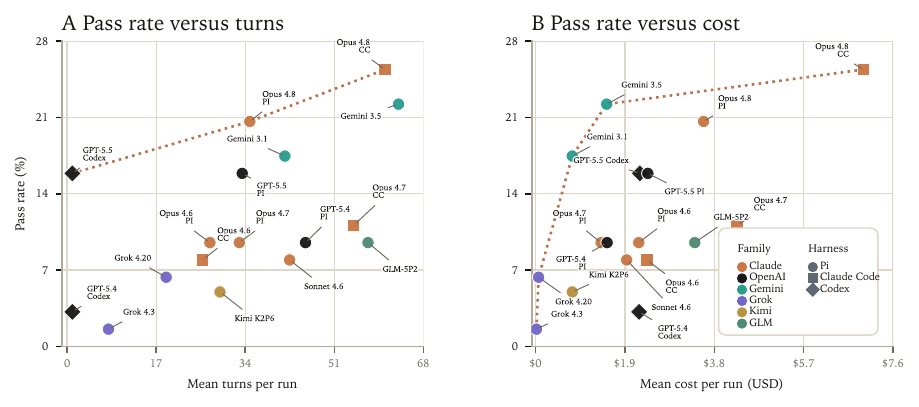}

  \captionof{figure}{\textbf{Efficiency frontiers for completed trajectories.}
  Pass rate versus mean turns and mean cost per completed run. Points are model--harness pairs;
  dotted lines mark Pareto frontiers. Turn metadata were complete; cost metadata were available
  for 1,050/1,068 trajectories.}
  \label{fig:topline-efficiency}
\end{center}

\StartBody

\section{Introduction}

Scientists use single-cell data to construct biological claims about cell
states, perturbations, development, immunity, and disease. Raw single-cell
measurements do not directly encode these claims. They must be processed
through multi-step workflows, interpreted against experimental design, and
integrated with assay-specific context, metadata, prior literature, and often
additional assay data such as immune repertoires, chromatin accessibility, or
functional validation experiments
\cite{lahnemann2020grand,heumos2023best,hao2021integrated}.

AI agents are beginning to show utility in biological data analysis, but
current benchmarks largely test broad biology knowledge without deep treatment
of the diverse task types specific to single-cell biology. Existing single-cell
benchmarks focus on local analysis steps and do not evaluate realistic
long-horizon science
\cite{laurent2024labbench,mitchener2025bixbench,nair2026compbiobench,li2026genebench,anthropic2026biomysterybench,workman2026scbench,workman2026spatialbenchlong}.

Verifiable ground truth is difficult to define in long-horizon single-cell
benchmarking. The same dataset can support multiple valid claims, and
published results do not always independently reproduce. A trustworthy
evaluation must therefore carefully constrain the set of admissible answers
without penalizing valid and potentially unanticipated analysis paths
\cite{ioannidis2005false,prinz2011believe,begley2012raise,errington2021replicability,workman2026spatialbenchlong,qu2026biomnibench,anthropic2026biomysterybench}.

We introduce scBench-Long, a benchmark for long-horizon single-cell biology.
scBench-Long contains 21 evaluations across human melanoma CD8 T-cell
reactivity, RNA+ATAC regulatory inference, human--monkey chimera development,
KRAS-driven lung tumor aging, and lethal COVID-19 lung pathology
\cite{ibanezmolero2026tumour,green2025enhancer,tan2021chimeric,shuldiner2025aging,melms2021molecular}.
Tasks provide raw or near-raw data, calibrated experimental context, auxiliary
resources when needed, and compact scientific questions. Agents must recover
structured biological conclusions that are graded deterministically over
controlled vocabularies and symbols, with companion rubrics used to diagnose
progress through key analysis chokepoints
\cite{workman2026scbench,workman2026spatialbenchlong,qu2026biomnibench}.

scBench-Long tests whether agents can move beyond accurately performing local
analysis steps to reasoning about the kinds of realistic scientific tasks found
in published studies or used in drug programs. We hope it serves as a useful
measuring device for evaluating frontier agent systems as they improve in
capability and are integrated into practical research and engineering.

\EndBody

\section{Benchmark Design}

{\fontsize{9.1}{12.8}\selectfont

Agents are evaluated on final scientific conclusions from 21 long-horizon
single-cell evaluations spanning five study systems: human melanoma
tumor-reactive CD8 T cells, RNA+ATAC regulation of CD8 tissue residency,
human--monkey chimeric embryo development, age-dependent KRAS-driven lung
tumorigenesis, and lethal COVID-19 lung pathology. The evaluations combine
single-cell and single-nucleus transcriptomes, paired TCR or BCR repertoires,
chromatin accessibility, cross-species ortholog mapping, ligand--receptor
databases, developmental references, and functional validation data
\cite{ibanezmolero2026tumour,green2025enhancer,tan2021chimeric,shuldiner2025aging,melms2021molecular}.\par}

{\fontsize{9.1}{12.8}\selectfont
\begin{center}
\begin{tikzpicture}
  \node[
    fill=figurecream,
    draw=figureline,
    line width=0.45pt,
    rounded corners=8pt,
    inner xsep=0.20in,
    inner ysep=0.14in,
    text width=0.91\textwidth,
    align=left
  ] {%
    \begin{minipage}{0.91\textwidth}
      {\fontsize{8.5}{10}\selectfont\bfseries\color{paperaccent}Benchmark Construction Workflow}\par
      \vspace{0.08in}
      \centering
      \includegraphics[width=\linewidth]{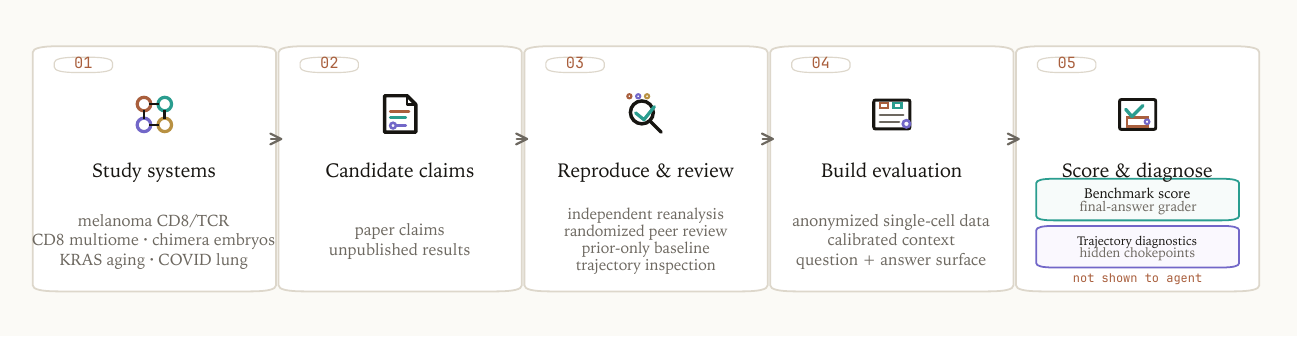}
      \captionof{figure}{\textbf{Benchmark construction workflow.}
      scBench-Long tasks are built by selecting single-cell study systems,
      calibrating candidate claims through reproduction and review, packaging
      anonymized data and controlled answer surfaces, and pairing
      deterministic final-answer grading with hidden trajectory diagnostics.}
      \label{fig:claim-conditioned-eval}
    \end{minipage}
  };
\end{tikzpicture}
\end{center}
}

{\fontsize{9.1}{12.8}\selectfont
\begin{multicols}{2}

scBench-Long is structured around biological studies rather than independent
datasets. Each study contributes multiple evaluations that revisit the same
experimental system from different scientific perspectives, requiring agents to
answer distinct questions about cell identity, state, regulation, composition,
and mechanism. This mirrors how single-cell datasets are analyzed in practice:
the same experiment is repeatedly interrogated to support different biological
claims \cite{lahnemann2020grand,heumos2023best}.

The benchmark builds on prior claim-conditioned benchmark construction rather
than introducing a new general grading framework. Each evaluation includes raw
or near-raw data, calibrated experimental context, scrubbed labels where needed,
a compact scientific question, a controlled final-answer surface, reproduction
notes, and trajectory rubrics. Paper claims are used as sources for candidate
evaluations, not as automatic ground truth. Candidate tasks are retained only
when the target conclusion can be reproduced from the staged data and hardened
through review, distractor design, and inspection of trajectories from multiple
model families
\cite{workman2026spatialbenchlong,workman2026scbench,nair2026compbiobench,qu2026biomnibench}.

The single-cell-specific contribution is the choice of study systems, evidence
layers, answer surfaces, and failure probes. The 21 evaluations span human
melanoma tumor-reactive CD8 T cell conjugates, RNA+ATAC regulation of
CD8 tissue-resident-like programs, human--monkey chimeric embryo development,
age-dependent KRAS-driven lung tumorigenesis, and lethal COVID-19 lung
pathology. The underlying data include 10x Chromium
5\textsuperscript{\ensuremath{\prime}} single-cell RNA sequencing with paired
TCR sequencing, 10x Multiome RNA and chromatin profiling, Smart-seq2
cross-species transcriptomics, Parse Biosciences combinatorial single-cell RNA
sequencing, and 10x Chromium 3\textsuperscript{\ensuremath{\prime}}
single-nucleus RNA sequencing. Several studies additionally require integrating
T or B cell receptor repertoires, chromatin accessibility, ortholog mappings,
ligand--receptor databases, developmental references, and
immunohistochemistry
\cite{zheng2017massively,han2014linking,picelli2014smartseq2,rosenberg2018splitseq,hao2021integrated,efremova2020cellphonedb,jin2021cellchat}.

Controlled answer surfaces are built around scientific objects that single-cell
analysts reason with: cell-state identities, donor compositions, regulator and
direction labels, clonotype or repertoire claims, ligand--receptor
interpretations, and mechanism-caution conclusions
\cite{workman2026spatialbenchlong,qu2026biomnibench}. Distractors are chosen to
separate data-supported conclusions from common shortcuts, including canonical
marker priors, raw-abundance rankings, single-modality answers, and causal
interpretations not supported by the data.

We use a small set of terms consistently throughout the paper. A \textit{final
evaluation} is one graded task with a controlled answer surface. A
\textit{replicate} is an independent attempt at that evaluation. A
\textit{trajectory} is the completed agent run and analysis trace for one
replicate. A \textit{model--harness pair} is a base model evaluated inside a
specific execution harness. The result set contains 17 model--harness pairs with
three replicates per final evaluation where runs completed, for 1,068 completed
trajectories.

\subsection{Verifiable grading paired with rubric diagnostics}

Benchmark scores use verifiable pass/fail grading on final scientific outcomes,
following the endpoint-grading strategy used in prior verifiable biology-agent
benchmarks
\cite{workman2026spatialbenchlong,workman2026scbench,nair2026compbiobench}.
Tasks require composing challenging local analysis steps: cell annotation,
donor statistics, repertoire analysis, chromatin inspection, reference mapping,
or literature-aware interpretation
\cite{lahnemann2020grand,heumos2023best,hao2021integrated}.

Verifiable endpoint grading is stable, but sparse. A model can fail the
benchmark while solving many subproblems correctly, and deterministic grading
necessarily penalizes answers outside the pre-specified target surface,
including some valid claims not anticipated by the benchmark authors
\cite{lightman2023verify,qu2026biomnibench}. We therefore retain
rubric-based trajectory judging as a companion diagnostic rather than as the
benchmark score. The general motivation for this design is shared with prior
long-horizon biological benchmarks; the scBench-Long rubrics focus on
single-cell chokepoints.

Evaluation authors define chokepoints after independent reproduction, review,
and inspection of trajectories from multiple model families
\cite{workman2026spatialbenchlong,qu2026biomnibench}. Chokepoint rubrics are
then used by LLM judges to score model trajectories, treating rubric scores as
diagnostic evidence rather than replacement benchmark scores
\cite{qu2026biomnibench,zheng2023llmjudge,wang2024fair,liu2023geval}.

\end{multicols}
}

\clearpage

\section{Evaluation Inventory}

{\fontsize{9.1}{12.8}\selectfont
scBench-Long is organized around study systems rather than independent
datasets. Each system contributes a cluster of evaluations that asks different
scientific questions with similar experimental context.

Across the five study systems, the benchmark contains 21 long-horizon
evaluations. The melanoma CD8 tasks require agents to link gene-expression
states with T-cell receptor clonotypes and interpret cell populations isolated
by marker-based sorting. The CD8 multiome tasks ask whether proposed regulatory
claims are supported by both RNA expression and chromatin accessibility. The
chimeric embryo tasks require comparison of developmental states across species
and evaluation of signaling between cell populations. The KRAS lung-tumor tasks
test whether agents can separate age-associated effects from tumor-cell state
and changes in the surrounding microenvironment. The COVID lung tasks require
integration of single-nucleus expression profiles with immune-receptor, cell
subset, and orthogonal validation evidence
\cite{ibanezmolero2026tumour,green2025enhancer,tan2021chimeric,shuldiner2025aging,melms2021molecular}.\par}

\begin{center}
\fontsize{7.2}{8.7}\selectfont
\setlength{\tabcolsep}{3.2pt}
\begin{tabular}{@{}>{\raggedright\arraybackslash}p{0.24\linewidth}
                >{\centering\arraybackslash}p{0.055\linewidth}
                >{\raggedright\arraybackslash}p{0.27\linewidth}
                >{\raggedright\arraybackslash}p{0.38\linewidth}@{}}
\toprule
Study/system & Evals & Assays / data types & Main task themes \\
\midrule
Melanoma tumor-reactive CD8 T cells &
5 &
10x 5\textsuperscript{\ensuremath{\prime}} scRNA-seq, paired scTCR-seq, sort-gate metadata &
Link gene-expression states with T-cell receptor clonotypes; interpret marker-sorted cell populations \\
\addlinespace
CD8 RNA+ATAC regulatory inference &
3 &
10x Multiome snRNA-seq and scATAC-seq &
Evaluate whether proposed regulatory claims are supported by both RNA expression and chromatin accessibility \\
\addlinespace
Human--monkey chimeric embryos &
4 &
Smart-seq2 scRNA-seq, ortholog maps, developmental references, ligand--receptor resources &
Compare developmental states across species; evaluate signaling between cell populations \\
\addlinespace
KRAS-driven lung tumor aging &
4 &
Parse Biosciences scRNA-seq, mouse lung tumor cohorts, external aging signatures &
Separate age-associated effects from tumor-cell state and surrounding microenvironment changes \\
\addlinespace
Lethal COVID-19 lung pathology &
5 &
Single-nucleus RNA-seq, BCR reconstruction, subset-level DE, immunohistochemistry &
Integrate single-nucleus expression profiles with immune-receptor, cell subset, and orthogonal validation evidence \\
\bottomrule
\end{tabular}
\captionof{table}{\textbf{Evaluation inventory.} scBench-Long groups evaluations by study
system while varying the claim, workflow, assay evidence, and controlled answer surface within each
system.}
\label{tab:evaluation-inventory}
\end{center}

\StartBody

\section{Results}

\subsection{Topline Benchmark Performance}

We aggregated 1,068 completed trajectories across 21 final evaluations and 17
model--harness pairs. Scores use deterministic pass/fail grading of the final
answer. Claude Opus 4.8 with Claude Code has the highest pass rate, with 16/63
passing trajectories (25.4\%; Wilson 95\% CI, 16.3--37.3) followed by Gemini
3.5 Flash and PI at 14/63 (22.2\%; 13.7--33.9).

Even the top agent systems solve only a minority of tasks consistently. Claude
Opus 4.8 with Claude Code passes at least one replicate for 8/21 evaluations,
at least two replicates for 6/21 evaluations, and all three replicates for 2/21
evaluations. Gemini 3.5 Flash with PI also reaches any-pass performance on 8/21
evaluations, but reaches majority-pass performance on 5/21 and all-replicate
performance on 1/21. Run-level pass rates therefore do not fully describe
reliability and replicate-level outcomes remain necessary for interpreting
long-horizon success.

Performance varies sharply across evaluations. Five evaluations have no
successful trajectories, while the highest-pass evaluation, a pseudotime-delay
task, has 26/51 passing trajectories (51.0\%). Most evaluations sit in low-pass
bins. Later results sections break these aggregate patterns into specific model
behaviors and failure modes.

Harness choice also changes performance. Claude Opus 4.8 is strongest with
Claude Code. GPT-5.4 is higher with PI than OpenAI Codex. GPT-5.5 has the same
run-level pass rate under PI and OpenAI Codex, but different majority-pass and
all-replicate counts.

\EndBody

\begin{center}
  \includegraphics[width=0.94\textwidth]{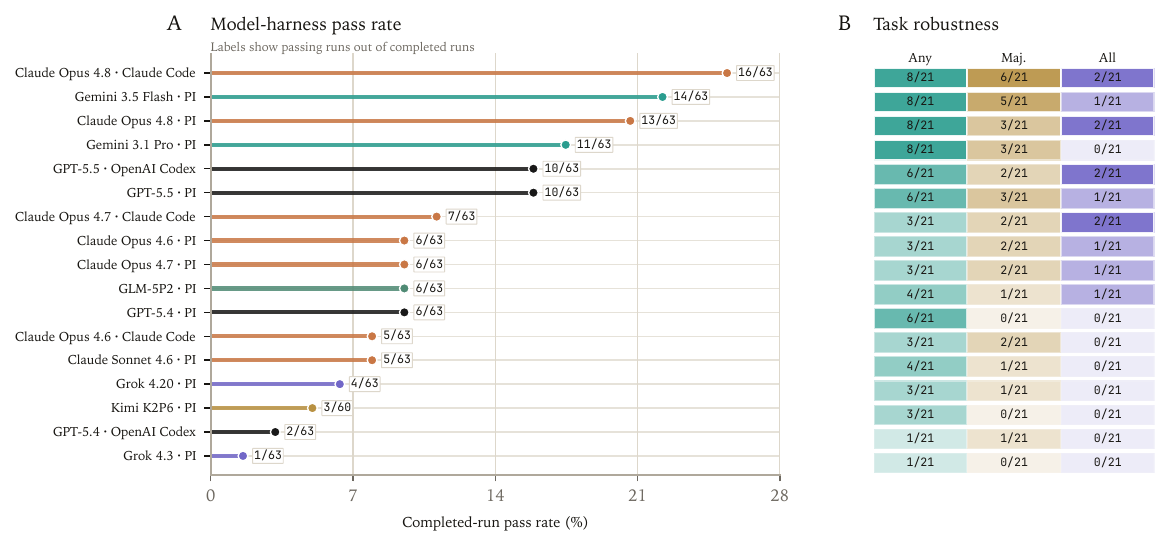}

  \captionof{figure}{\textbf{Main results summary.} Completed-run pass rates and task-level
  robustness counts across 21 final evaluations and 17 model--harness pairs. Panel A reports passing
  runs out of completed runs. Panel B reports the number of evaluations with any passing replicate,
  a majority of passing replicates, or all replicates passing. Wilson confidence intervals are
  reported in Table~\ref{tab:model-results}.}
  \label{fig:main-results-summary}
\end{center}

\vspace{0.08in}

\begin{center}
{\fontsize{5.5}{6.8}\selectfont
\begin{tabular}{@{}lrrrrrr@{}}
\toprule
Model--harness & Pass runs & Pass \% & Wilson 95\% CI & Any & Majority & All \\
\midrule
Claude Opus 4.8 / Claude Code & 16/63 & 25.40 & 16.28--37.34 & 8/21 & 6/21 & 2/21 \\
Gemini 3.5 Flash / PI & 14/63 & 22.22 & 13.72--33.91 & 8/21 & 5/21 & 1/21 \\
Claude Opus 4.8 / PI & 13/63 & 20.63 & 12.48--32.17 & 8/21 & 3/21 & 2/21 \\
Gemini 3.1 Pro / PI & 11/63 & 17.46 & 10.04--28.62 & 8/21 & 3/21 & 0/21 \\
GPT-5.5 / OpenAI Codex & 10/63 & 15.87 & 8.86--26.81 & 6/21 & 2/21 & 2/21 \\
GPT-5.5 / PI & 10/63 & 15.87 & 8.86--26.81 & 6/21 & 3/21 & 1/21 \\
Claude Opus 4.7 / Claude Code & 7/63 & 11.11 & 5.49--21.20 & 3/21 & 2/21 & 2/21 \\
Claude Opus 4.6 / PI & 6/63 & 9.52 & 4.44--19.26 & 3/21 & 2/21 & 1/21 \\
Claude Opus 4.7 / PI & 6/63 & 9.52 & 4.44--19.26 & 3/21 & 2/21 & 1/21 \\
GLM-5P2 / PI & 6/63 & 9.52 & 4.44--19.26 & 4/21 & 1/21 & 1/21 \\
GPT-5.4 / PI & 6/63 & 9.52 & 4.44--19.26 & 6/21 & 0/21 & 0/21 \\
Claude Opus 4.6 / Claude Code & 5/63 & 7.94 & 3.44--17.27 & 3/21 & 2/21 & 0/21 \\
Claude Sonnet 4.6 / PI & 5/63 & 7.94 & 3.44--17.27 & 4/21 & 1/21 & 0/21 \\
Grok 4.20 / PI & 4/63 & 6.35 & 2.50--15.22 & 3/21 & 1/21 & 0/21 \\
Kimi K2P6 / PI & 3/60 & 5.00 & 1.72--13.70 & 3/21 & 0/21 & 0/21 \\
GPT-5.4 / OpenAI Codex & 2/63 & 3.17 & 0.88--10.86 & 1/21 & 1/21 & 0/21 \\
Grok 4.3 / PI & 1/63 & 1.59 & 0.28--8.46 & 1/21 & 0/21 & 0/21 \\
\bottomrule
\end{tabular}
}

\captionof{table}{\textbf{Main verifiable results.} Model--harness pass rates across the 21 final
evaluations. Wilson intervals summarize run-level uncertainty. Any, Majority, and All report the
number of evaluations with at least one passing replicate, a majority of passing replicates, or all
replicates passing, respectively.}
\label{tab:model-results}
\end{center}

\vspace{0.08in}
\fontsize{9.1}{12.8}\selectfont
\begin{multicols}{2}

\subsection{Rubric Judges Provide Dense but Imperfect Trajectory Diagnostics}

Endpoint pass/fail grading, the deterministic final-answer score, remains the benchmark score, but it is too sparse to
describe partial scientific progress. We therefore scored trajectories with four
rubric judges using task-specific chokepoints. In the GLM-inclusive judge
matrix, 1,067/1,068 trajectories had at least one judge row, 1,061 had at least
three judge scores, 855 had all four judge scores, and 4,050/4,272 expected
judge-score cells were present. Endpoint labels below come from the deterministic
main results manifest; two stale fourth-replicate judge rows were excluded.

\subsubsection{Judge scores are reproducible}

Judge scores were consistent enough for diagnostic use. Mean pairwise judge
correlation was 0.90, the minimum pairwise correlation was 0.87, and the mean
absolute pairwise score difference was 7.6 percentage points. Same-trajectory
judge variability was similar to variation across independent attempts on the
same evaluation and model--harness group: judge SD averaged 5.5 points per
trajectory, while replicate SD averaged 6.3 points.

\subsubsection{Rubric scores enrich for success}

Endpoint success remained sparse, with 125/1,068 passing trajectories (11.7\%).
By contrast, mean rubric score was nonzero for 1,066/1,068 trajectories. Among
trajectories with at least three judge scores, endpoint-passing runs had higher
mean rubric scores than endpoint-failing runs (75.9\% vs. 58.7\%). The
association was useful but incomplete: Pearson $r=0.29$, Spearman
$\rho=0.30$, and AUC $=0.77$. The highest rubric decile contained 47/106
endpoint passes, so high rubric scores enrich for success but do not replace
deterministic grading.

\subsubsection{Rubric scores capture trajectory style}

Rubric scores also varied by source model--harness. Opus 4.8 and GPT-5.5
trajectories received high rubric scores, GLM was intermediate, and Grok runs
were lower. This ordering was not identical to endpoint pass rate: for example,
Opus 4.8 with PI had a slightly higher mean rubric score than Opus 4.8 with
Claude Code, while Claude Code had the higher endpoint pass rate. We therefore
treat rubric scores as trajectory-style annotations and partial-progress
diagnostics, not as calibrated scientific correctness scores.

\end{multicols}

\begin{center}
  \includegraphics[width=\textwidth]{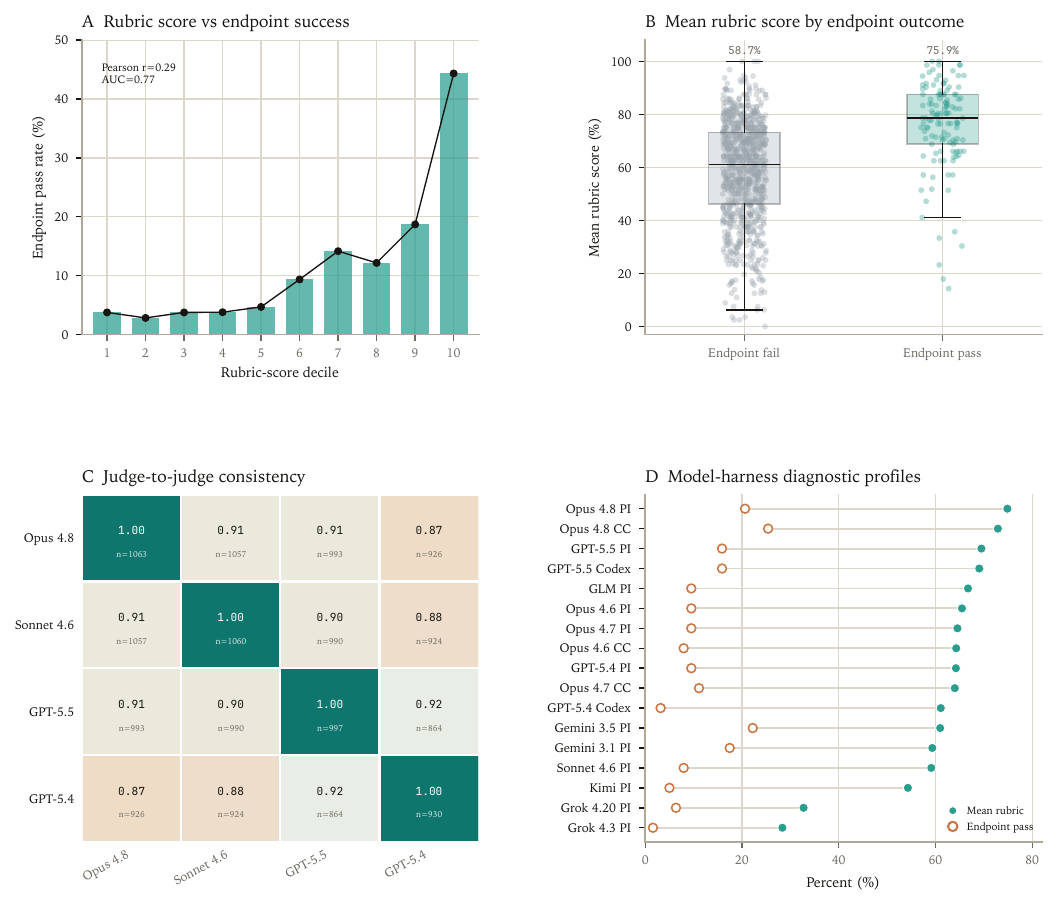}

  \captionof{figure}{\textbf{Rubric judges provide dense but imperfect diagnostics.}
  Panel A shows endpoint pass rate by rubric-score decile, with endpoint correlation and AUC
  reported as diagnostics. Panel B shows mean rubric score by deterministic endpoint outcome.
  Panel C reports pairwise judge correlations on matched scored trajectories. Panel D compares
  endpoint pass rate and mean rubric score within each model--harness group. Endpoint labels use
  the GLM-inclusive deterministic results manifest; the single missing judge row is counted in the
  endpoint denominator as a fail and omitted from rubric-score means.}
  \label{fig:rubric-judge-diagnostics}
\end{center}

\fontsize{9.1}{12.8}\selectfont
\begin{multicols}{2}

\subsection{Failure Modes in Long-Horizon Biological Reasoning}

Across the benchmark, models often recovered individual components of an analysis without
assembling them into the correct biological conclusion. Claude, GPT, Gemini, GLM, Grok, and Kimi
frequently identified relevant cell populations, recovered plausible ligand-receptor interactions,
reconstructed immune receptor repertoires, and inferred broad regulatory programs. These
intermediate results did not reliably translate into correct scientific claims. The recurring
failure modes were not simple tool-use failures. Models prioritized abundant signals over specific
evidence, substituted canonical biological expectations for measurements, drew causal conclusions
from cross-sectional data, and simplified regulatory relationships that required integrating
multiple modalities.

\subsubsection{\texorpdfstring{Familiar Biological Priors Can Mask\\ Task-Specific Evidence}{Familiar Biological Priors Can Mask Task-Specific Evidence}}

Models will sometimes submit an answer based on a familiar biological prior,
often some plausible conclusion from existing literature that shares some of the
decision context, even after completing many correct analysis steps.

For instance, in melanoma CD8 T-cell tasks, models leaned on the familiar
association between CD39, PD-1, TOX, LAG3, HAVCR2, CXCL13, and exhausted,
antigen-experienced tumor-infiltrating CD8 T cells. This led them to treat
CD39-high or PD-1-high singlets as the most tumor-reactive population, even
though the task required combining physical tumor/APC engagement with paired TCR
clonal expansion.

Another task modified the input data by swapping two metadata labels,
\texttt{singlet} and \texttt{tumor\_cluster}, while leaving the expression
matrix and TCR data unchanged. This was to test whether an agent investigated
the structure of the data empirically without drawing quick decisions from
labels. Many models either relabeled the result back to the \mbox{expected biology} or
partially followed the data and then overgeneralized.

\subsubsection{\texorpdfstring{Raw abundance can be mistaken\\ for biological importance}{Raw abundance can be mistaken for biological importance}}

Models often overattributed scientific importance to large quantitative values
in several tasks.

In a human--monkey chimera task, agents analyzed ligand--receptor communication
between hEPSC-derived human epiblast cells and host monkey epiblast cells from
the same embryos \cite{tan2021chimeric,efremova2020cellphonedb,jin2021cellchat}.
The task required agents to decide whether significant
ligand--receptor pairs supported a single dominant signaling family after
accounting for directionality and database composition. Agents often treated
the largest raw interaction class as the answer. However, the data contained
many protein ligand--receptor pairs in both directions, with no single family
dominating robustly in aggregate.

Only 2/50 trajectories passed, and both GPT-5.5 and Opus 4.8 were 0/6 across
their paired harnesses.

\subsubsection{\texorpdfstring{Association Can Be Mistaken\\ for Mechanism}{Association Can Be Mistaken for Mechanism}}

In the melanoma tumor-conjugate task, agents compared malignant cells recovered
from tumor:T-cell conjugates with malignant singlets. The task was to identify
which tumor-cell programs were enriched in the conjugate population and to
interpret those enrichments cautiously. Because the cells were sampled after
physical pairing, the data show that certain tumor programs co-occur with T-cell
contact. They do not, by themselves, show whether those programs were present
before contact, induced by contact, or partly contaminated by transcripts from
the co-sorted T cell.

The data supported two enriched malignant-cell programs with different levels of
interpretability. A hypoxia/HIF program was robust across patients and controls,
consistent with a tumor-intrinsic state enriched among conjugate-associated
malignant cells. By contrast, the immune-response program was more sensitive to
decontamination and could not be assigned a direction from these data alone. It
could reflect tumor cells predisposed to engage T cells, tumor cells responding
to contact, selection of particular tumor states into conjugates, or residual
RNA from the paired lymphocyte.

No model--harness pair passed this evaluation. Models often recovered plausible
tumor states, interferon ligands, or antigen-presentation signals, but treated
enrichment in the conjugate-associated population as evidence of a directional
mechanism.

\subsubsection{Failing to integrate multiple modalities}

CD8 multiome tasks asked agents to infer regulatory control of the
CD103+ Trm-like TIL program from paired snRNA-seq and scATAC-seq. These tasks
required recognizing molecular patterns across RNA and chromatin accessibility
\cite{green2025enhancer,hao2021integrated}.

Model failures often only considered a single modality. In the KLF2 tasks,
models often followed RNA differential expression and nominated genes such as
\textit{S1pr5} or \textit{Cd160}, even though the relevant enhancer data did
not support them as regulators of the Trm-like program. In another case, models
recovered that KLF2 behaved as a repressor, but inferred its activity from
\textit{Klf2} RNA rather than from the coordinated behavior of its targets.

\subsection{Harness effects in paired model comparisons}

Harness choice changed endpoint outcomes for the same underlying model. We
restricted this analysis to models evaluated under two harnesses: Claude Opus
4.6--4.8 under Claude Code and PI, and GPT-5.4--5.5 under OpenAI Codex and PI.
Aggregate pass rate shifted for four of the five paired models. Claude Opus 4.8
was higher with Claude Code than PI (16/63 vs. 13/63), Claude Opus 4.7 was
slightly higher with Claude Code (7/63 vs. 6/63), Claude Opus 4.6 was slightly
higher with PI (6/63 vs. 5/63), and GPT-5.4 was higher with PI (6/63 vs. 2/63).
GPT-5.5 had the same aggregate pass rate under OpenAI Codex and PI (10/63 for
both).

Aggregate pass rate understated the harness effect. For each paired model, 3--6
of 21 evaluations changed pass count by harness, and GPT-5.5 solved different
evaluations despite equal overall pass rate. These shifts show that
long-horizon single-cell performance depends on the model--harness system, not
only on the base model. We therefore report results at the model--harness level
and interpret harnesses as part of the evaluated agent.

\EndBody

\begin{center}
  \includegraphics[width=\textwidth]{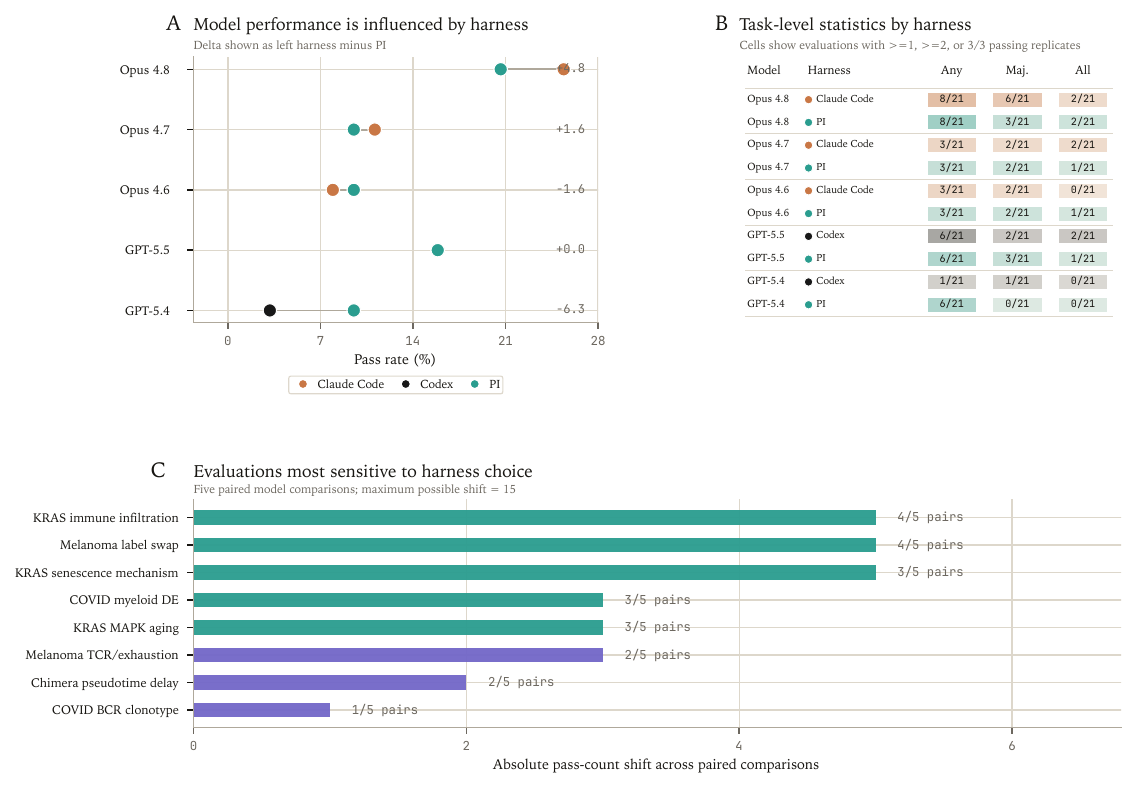}
  \captionof{figure}{\textbf{Harness effects in paired model comparisons.}
  Panel A compares completed-run pass rates for identical models evaluated under two harnesses.
  Delta labels report the non-PI harness minus PI. Panel B reports task-level statistics by
  model--harness group: cells show evaluations with at least one, at least two, or all three passing
  replicates. Panel C ranks evaluations by the total absolute shift in passed replicates across
  paired harness comparisons.}
  \label{fig:harness-effects}
\end{center}

\StartBody

\section{Discussion}

scBench-Long evaluates whether AI agents can recover claim-conditioned
scientific conclusions from raw or near-raw single-cell data. The benchmark
stages the data and context needed to test a specific biological claim, then
grades the final structured answer deterministically. This follows prior
verifiable long-horizon benchmark design while adapting it to single-cell tasks
that require donor-aware reasoning, immune-repertoire analysis, chromatin
integration, cross-species mapping, and orthogonal validation evidence
\cite{workman2026spatialbenchlong,workman2026scbench,qu2026biomnibench}.

Current agents show limited but measurable capability. The strongest
model--harness pair, Claude Opus 4.8 with Claude Code, passes 16/63 completed
runs (25.4\%), while Gemini 3.5 Flash with PI passes 14/63 (22.2\%). Success is
not yet reliable: the top pair passes at least one replicate for 8/21
evaluations and all three replicates for only 2/21.

Models often execute plausible single-cell workflows but fail to reach the
correct biological conclusion. We observed several recurring failure modes:
substituting prior literature expectations for empirical evidence, conflating
large numerical values with scientific importance, inferring mechanisms from
molecular associations, and failing to integrate evidence across modalities.
scBench-Long therefore evaluates agents beyond their ability to run local
analysis steps, testing whether they can complete multi-step analyses and
decide which final claims are supported by the data.

We hope scBench-Long serves both as a measurement tool and a diagnostic lens for
developing agents that analyze single-cell data faithfully, transparently, and
reproducibly. It is a focused contribution within a broader benchmark family
spanning major biological data classes and work categories, including spatial
biology, epigenomics, and therapeutics. More broadly, we view benchmarks as
evolving specifications of computational biology workflows, supporting
test-driven development of agent systems whose behavior can improve through
both model training and harness engineering.

\columnbreak

\section{Methods}

\subsection{Evaluation construction}

scBench-Long evaluations were constructed from study systems in which a
specific single-cell claim could be tested from staged data and calibrated
experimental context. We selected candidate claims from source studies and
internal reproduction notes, then retained tasks only when the target
conclusion could be independently reproduced from the data available to the
agent. Source papers were therefore used as claim generators rather than as
automatic ground truth.

Each evaluation was packaged as a task specification containing a compact
scientific question, raw or near-raw data nodes, any required auxiliary
resources, a controlled answer surface, metadata describing the biological
dependency stack, and a deterministic grader. Data nodes included single-cell
or single-nucleus expression matrices, immune-receptor tables, chromatin
accessibility or multiome objects, cross-species mapping resources,
ligand--receptor resources, developmental references, and vocabulary files as
needed by the task. Task prompts were written to provide the experimental
context needed for analysis without prescribing a solution path.

\subsection{Constructing ground truths}

Final answers were structured JSON objects over controlled vocabularies and
symbols. Answer surfaces were chosen to match the scientific objects used by
single-cell analysts, including cell-state labels, donor-level composition
calls, regulator identities, direction labels, clonotype claims,
ligand--receptor interpretations, and explicit mechanism-caution labels.
Distractors and hard-fail conditions were added during review to separate
data-supported conclusions from common shortcuts such as canonical marker
recall, abundance-only ranking, single-modality reasoning, or causal
interpretation of cross-sectional associations.

Benchmark scores use deterministic pass/fail grading of the final answer. Each
grader combines typed checks over the submitted JSON answer, including list
matching, dictionary matching, field-level predicates, required keys,
multi-accepted synonyms or resolution-sensitive labels, and hard-fail
conditions for biologically incompatible answers. Grading is intentionally
endpoint-based: intermediate analyses are not required to follow a prescribed
workflow, and agents can pass through any valid analysis path that reaches the
controlled target surface.

\subsection{Model-harness runs and result aggregation}

The analyzed result set contains 21 final evaluations and 17 model--harness
pairs. Each model--harness pair was run for three replicates per evaluation
where runs completed, yielding 1,068 completed trajectories. The evaluated
harnesses were PI, Claude Code, and OpenAI Codex. The evaluated model families
included Claude, Gemini, OpenAI GPT, Grok, Kimi, and GLM systems.

Run-level results were aggregated from the GLM-inclusive results manifest
\texttt{results/scbench-long-results-w-glm.csv}. Each row records the model, harness,
evaluation identifier, replicate, remote paths to the evaluation and result
artifacts, and deterministic final-pass status. Model-harness summaries report
completed runs, passing runs, Wilson confidence intervals, and task-level
robustness counts: evaluations with at least one passing replicate, a majority
of passing replicates, or all completed replicates passing. Missing attempts
were excluded from run denominators and tracked separately during aggregation.

Cost and turn metadata were extracted from per-run result JSON artifacts when
available. Turn metadata were complete for the analyzed trajectories; cost
metadata were available for 1,050/1,068 trajectories. Efficiency summaries use
mean cost and mean turn count over completed runs and are interpreted as
harness- and logging-dependent diagnostics rather than benchmark scores.

\subsection{Rubric-based trajectory diagnostics}

Endpoint grading is stable but sparse, so each evaluation also includes
trajectory rubrics for diagnostic analysis. Rubrics were written after
reproduction and review to capture single-cell chokepoints expected to matter
across plausible solution paths, such as identifying the correct biological
comparison, preserving donor structure, using receptor evidence appropriately,
checking chromatin support for RNA-level claims, mapping cross-species states,
or avoiding causal claims unsupported by the data.

Rubric judges scored model trajectories independently of the deterministic
benchmark score. The GLM-inclusive judge matrix contains one row per judged
trajectory and per-judge fractional rubric scores from four judge models when
available. Rubric summaries use matched deterministic endpoint labels from the
main results manifest. Because rubric scores are prompt- and judge-dependent,
they are treated as companion diagnostics for partial progress and failure-mode
analysis, not as replacement benchmark scores.

\EndBody

\end{document}